\documentstyle[aps]{revtex}
\def\beq{\begin{eqnarray}}
\def\eed{\end{eqnarray}}

\begin{document}
\tightenlines
\draft
\title{Spin-Liquid State for Two-Dimensional Heisenberg Antiferromagnets on a
Triangular Lattice}
\author{Zhan-hai Dong$^1$ and Shiping Feng$^{1,2,3}$ }
\address{$^{1*}$Department of Physics, Beijing Normal University, Beijing
 100875, China  \\
$^{2}$National Laboratory of Superconductivity, Academia Sinica,
 Beijing 100080, China \\
$^{3}$Institute of Theoretical Physics, Academia Sinica,
 Beijing 100080, China \\}
\maketitle
\begin{abstract}
The spin liquid state of the antiferromagnetic Heisenberg model on a
triangular lattice is studied within the self-consistent Green's function
method. It is shown that the spin excitation spectra is gapless, and
ground-state energy per site is $E_{g}/NJ=-0.966$, which is in very good
agreement with the results obtained within the variational Monte Carlo
method based on the resonating-valence-bond state. Some thermodynamic
properties are also discussed.
\end{abstract}
\pacs{75.10.Jm, 67.40.Db, 75.10.+x}

In the last decades, a large amount of work has been devoted to the
understanding of the property of the two-dimensional (2D) quantum Heisenberg
antiferromagnets. This is motivated partly by the search for quantum
disordered ground-state and their possible relationship to the high
temperature superconductivity \cite{1}. The numerical simulations strongly
support the existence of the antiferromagnetic (AF) long-range-order (AFLRO)
for the Heisenberg antiferromagnet on a square lattice with the reduced
magnetic moment of about $60\%$ of its classical value \cite{2}. It has been
shown that this magnetic moment is reduced further by considering the spin
frustration arising from the next-nearest-neighbors coupling \cite{3}. In
contrast with the Heisenberg antiferromagnet on the square lattice, the
Heisenberg antiferromagnet on a triangular lattice is the three-sublattice
spin system with geometric frustration. Anderson \cite{4} argued that the
strong spin frustration on the triangular lattice may generate a novel
resonating-valence-bond (RVB) spin liquid ground-state without AFLRO. Since
then some different types of quantum ground-state have been proposed, in
particular, Kalmeyer and Laughlin \cite{5} argued that RVB state for the
triangular lattice is similar to the fractional quantum Hall state for
bosons. Moreover, various numerical techniques and analytical approaches
have been used to study this system \cite{6,7}. The variational energy
reported by Lee and Feng \cite{8} was essentially identical to the result
obtained by Anderson \cite{4} based on the RVB state. Huse and Elser \cite{9}
subsequently constructed wave functions exhibiting AFLRO which reportedly had
a lower variational energy than Anderson's \cite{4}. Therefore the nature of
the ground-state, especially the existence of AFLRO for the AF Heisenberg
antiferromagnet on the triangular lattice, is still controversial.

Apart from the numerical techniques, a popular method to study the quantum
spin problem is the spin -wave theory, where the quantum spin is mapped onto
the boson representation in terms of the Holstein-Primakoff transformation.
However, the native spin-wave theory, an expansion and linear approximation
in Holstein-Primakoff bosons, leads to violation of the commutation rule of
the quantum spin, this is because that the quantum spins obey the Pauli
algebra, i.e., the spin one-half raising and lowering operators behave as
fermions on the same site and as bosons on different sites. In this paper,
we study the spin liquid state of the AF Heisenberg antiferromagnet
on the triangular lattice within the Green's function theory under the
Kondo-Yamaji \cite{10} decoupling scheme. With the help of the spectral
representations of the correlations, the self-consistent equations are
obtained to determine the order parameters. The spin excitation spectral is
gapless and spin liquid ground-state energy per site is $E_g/NJ=-0.966$.
Some thermodynamic properties are also discussed.

The standard AF Heisenberg model on the triangular lattice is written as, 
\begin{eqnarray}
H=J\sum_{i\eta }{\bf S}_i \cdot {\bf S}_{i+\eta },
\end{eqnarray}
where the sum is over all sites $i$ and, for each $i$, over
nearest-neighbors $\eta $. Since the quantum spin operators obey the Pauli
spin algebra, then it can be discussed in terms of the Tyablikov's two-time
Green's function method \cite{11}. We define the spin two-time Green's
function $G$ as
\begin{eqnarray}
G(i-j, t-t^{\prime })=-i \theta (t-t^{\prime })<[S_i^{+}(t),S_j^{-}
(t^{\prime })]> \equiv <<S_i^{+}(t); S_j^{-}(t^{\prime })>>,
\end{eqnarray}
with $\theta (t)$ is the step function, $S_i^{+}$ and $S_i^{-}$ are the
raising and lowering operators of $S_i^z$, respectively, and $<\cdots >$ is
an average over the ensemble. The time-Fourier transform of the two-time
Green's function satisfies the equation, \begin{eqnarray}
\omega <<A; B>>=\frac 1{2\pi }<[A,B]>+<<[A,H];B>>, 
\end{eqnarray}
and the correlation functions can be obtained by the spectral
representations as,
\begin{eqnarray}
<B(t^{\prime })A(t)>=i \int_{-\infty }^\infty \frac{{\rm d}\omega }{2\pi 
} \frac{<<A;B>>_{\omega +i0^{+}}-<<A;B>>_{\omega -i0^{+}}}{e^{\beta \omega
}-1} e{^{{-i\omega (t-t^{\prime })}}}.
\end{eqnarray}
With the help of the Eq. (3), the equation of motion of the spin Green's
function (3) is evaluated as,
\begin{eqnarray}
\omega G(i-j, \omega )=\delta _{ij}<S_i^z> +2J\sum_\eta <<S_{i+\eta
}^{+}S_i^z- S_i^{+}S_{i+\eta }^z; S_j^{-}>>_\omega.
\end{eqnarray}
Since the three-sublattice ordering for the triangular lattice is still
controversial, and the spin liquid hypothesis is not convincingly discarded,
then in the following discussions, we only study the system in the case
$<S_i^z>=0$, i.e., the spin liquid state without AFLRO. In this case, the
basic equation for the spin Green's function in the one-dimension have been
discussed in detail by Kondo and Yamaji \cite{10}, and 2D by many authors
\cite{12}. Following their discussions, we can obtain the spin Green's
function (5) as,
\begin{eqnarray}
G(k,\omega )=\frac{ZJC_1}\pi \frac{\gamma _k-1}{\omega ^2-\omega ^2(k)},
\end{eqnarray}
where $Z$ is the number of the nearest neighbor sites,
$\gamma_k=(1/Z)\sum_\eta e^{ik\cdot \eta}$, and spin excitation spectra,
\begin{eqnarray}
\omega ^2(k) =ZJ^2[\frac 12+\alpha C_2-\alpha C_1(1+Z\gamma _k)]
(1-\gamma_k),
\end{eqnarray}
with the order parameters $C_1=<S_i^{+}S_{i+\eta }^{-}>$, $C_2=\sum_{\eta
\eta^{\prime }(\eta \neq \eta ^{\prime })}<S_{i+\eta }^{+}S{_{i+
\eta ^{\prime}}^{-}}>$. In order not to violate the sum rule of the
correlation function $<S_i^{+}S_i^{-}>=1/2$ in the case $<S_i^z>=0$,
the important decoupling parameter $\alpha $ has been introduced as
these discussed by Kondo and Yamaji \cite{10}, which is regarded as the
vertex corrections. With the help
of the Green's function (6) and the spectra representation of the
correlation functions (4), we obtain the self-consistent equations to
determine the order parameters $C_1,C_2$, and $\alpha $ as, 
\begin{eqnarray}
C_1&=&-ZJC_1 \frac1N \sum_k \gamma_k \frac {1-\gamma_k}{\omega(k)}
\coth [\frac12 \beta \omega(k)],\\
C_2&=&-ZJC_1 \frac1N \sum_k (Z\gamma_k^2-1) \frac {1-\gamma_k}{\omega(k)}
\coth [\frac12 \beta \omega(k)],\\
\frac12&=&-ZJC_1 \frac1N \sum_k \frac {1-\gamma_k}{\omega(k)}
\coth [\frac12 \beta \omega(k)].
\end{eqnarray}

We have performed the numerical calculation for the above self-consistent
equations. The ground-state energy per site is $E_g/NJ=-0.966$. For
comparison with the other results obtained by the numerical simulations, the
present theoretical result is also shown in Table I. It is obvious that our
result of the ground-state energy is in very good agreement with those
obtained by Lee and Feng \cite{8} based on the variational d-wave RVB state
without AFLRO, but higher than the results of Huse and Elser \cite{9} for
a AFLRO state and Nishimori and Nakanishi's \cite {6} finite lattice exact
diagonalization. Huse and Elser \cite{9} have estimated that the
ground-state energy per site is $E_g/NJ\le -1.074$, then
our ground-state energy is more than $10\%$ higher this estimates. But the
present low energy of the spin liquid state and the short range of its spin
correlations indicate that longer-range spin interactions, e.g.,
second-near-neighbor coupling, can push the system further into the liquid
phase. More interestingly, the excitation spectra of the RVB state for the
triangular lattice obtained by Kalmeyer and Laughlin \cite{5} or Lee and
Feng \cite{8} has a large gap, but in the present result, the excitation
spectra of the spin liquid state is gapless, which is shown in Fig. 1, and
the minimum-energy excitation is the spinon at $k_c=(0,0)$. Our spin
liquid state has spin-${1\over 2}$ bosonic spinons, and this structure
clearly disagree with the Kalmeyer and Laughlin \cite{5,15} state on the
triangular lattice, which was argued to possess semionic spinons. We have
also computed the spin structure factor
$S(k)=(1/N)\sum_{<ij>}<S_i^z S_j^z> e^{ik\cdot (R_{i}-R_{j})}$
in the zero temperature, and the result is plotted
in Fig. 2, the maxima of $S(k)$ occur at six wave vectors, which is
consistent with the result obtained within the large-N expansion
\cite{14}.

Some interesting properties associated the present spin liquid state, such
as, the specific heat and susceptibility $\chi=(g^2 \mu_B^2/k_BT)
\sum_{ij}<S_i^zS_j^z>$ with $g$ is the Landa factor and $\mu_B$ is the Bohr
magneton, have been discussed, and the results are shown in Fig. 3(a) and
Fig. 3(b), respectively. The specific heat display a broad peak near $k_BT
\sim 0.5J$, which is in qualitative agreement with results of the AF
Heisenberg model on the square lattice \cite{13}. The susceptibility is very
weakly dependent on $T$, and the similar behavior has been obtained on the
square lattice \cite{13}. Therefore, our present results seem to indicate
that the global behavior of the AF Heisenberg model on the triangular
lattice, which is the system with geometric frustration, is qualitative
consistent with the frustrated AF Heisenberg model on the square lattice.

In summary, we have studied the spin liquid state of the AF Heisenberg model
on the triangular lattice within the Kondo-Yamaji's decoupling scheme. It is
shown that the spin excitation spectra is gapless, and the ground-state
energy per site is $E_g/NJ=-0.966$. The energy of this spin liquid is very
good consistent with the RVB state, but higher than the result for the AFLRO
state. The behaviors of the specific heat and susceptibility are similar
to these obtained on the square lattice.

A natural question is what is the reason why this self-consistent mean-field
Green's function theory is so useful to treat the quantum spin systems
without AFLRO? To our present understanding, there are at least two reasons:
(1) the commutation rule (Pauli algebra) of the quantum spin is exactly
satisfied in the actual calculations. (2) The sum rule of the spin Green's
function is always satisfied, and the rotational symmetry is not
unphysically broken in this self-consistent mean-field Green's function
theory. For the spin systems without AFLRO, the low lying excitations
described are essentially spin waves propagating in a short-range-order
with a correlation length. Kondo and Yamaji \cite{10},  and Shimahara and
Takada \cite{12} have employed this self-consistent mean-field Green's
function theory to study the one-dimensional Heisenberg spin system and
2D AF Heisenberg spin system on the square lattice, respectively, they
obtained the results which have satisfactory temperature dependence over
whole temperature region from qualitative view point.

\acknowledgments
The authors would like to thank Yun Song and Zhongbing Huang for helpful
discussions. This work was supported by the National Natural Science
Foundation Grant No. 19774014 and the Trans-Century Training Programme
Foundation for the Talents by the State Education Commission of China.

\begin{figure}
\caption{Momentum dependence of the spin excitation spectra $\omega (k)$
of the spin liquid state. The minimum-energy excitation is the spinon at
$k_c=(0,0)$}
\label{autonum}
\end{figure}

\begin{figure}
\caption{Zero temperature structure factor S(k) of the spin liquid state.
The global maxima are at six wave vectors.}
\label{autonum}
\end{figure}

\begin{figure}
\caption{(a) The specific heat and (b) susceptibility as a function of
temperature.}
\label{autonum}
\end{figure}

\begin{table}
\caption{A comparison of the ground-state energy per site for the 
antiferromagnetic Heisenberg model on the two-dimensional triangular lattice}
\begin{tabular}{l|l|l}
Author(s)           &   $E_g/NJ$  &   Method \\
\hline
Kalmeyer and Laughlin  & $-0.94(2)$  &  Variational RVB \\
\hline
Huse and Elser  & $-1.02(2)$  &  Variational AF \\
\hline
Nishimori and Nakanishi & $-1.10(2)$ & Finite lattice \\
Lee and Feng & $-0.968(4)$ & Variational RVB \\
\hline
The present work & $-0.966$ & Green's function method \\
\end{tabular}
\end{table}

\end{document}